%% file: qdislib.tex
\documentclass[10pt,conference]{IEEEtran}

\AtBeginDocument{%
  }
    
\usepackage{algorithmic}
\usepackage{graphicx}
\usepackage{textcomp}
\usepackage{xcolor}
\usepackage{listings}
\usepackage{minted}
\usepackage{todonotes}
\usepackage{fancyhdr} 

\usepackage{listings}
\usepackage{xcolor}
\usepackage{ dsfont }
\usepackage[inkscapelatex=false]{svg}

\usepackage{tikz}
\usetikzlibrary{quantikz2}
\usepackage{tikzsymbols}

\usepackage[most]{tcolorbox}
\usepackage{booktabs}
\usepackage{caption}

\usepackage{comment}
\usepackage{enumitem}

\usepackage{graphicx}
\usepackage{subcaption}
\usepackage{multirow}
\usepackage{hyperref}

\usepackage{pgfplots}
\pgfplotsset{compat=1.18}
\usepackage{booktabs}
\usepackage{multirow}      
\usepackage{pgfplotstable} 
\usepackage{xcolor}

\definecolor{codegreen}{rgb}{0,0.6,0}
\definecolor{codegray}{rgb}{0.5,0.5,0.5}
\definecolor{codepurple}{rgb}{0.58,0,0.82}
\definecolor{backcolour}{rgb}{0.95,0.95,0.92}
\definecolor{backcode}{RGB}{252,251,220}
\definecolor{bluebsc}{RGB}{31,57,112}

\lstdefinestyle{mystyle}{
    backgroundcolor=\color{backcode},   
    showstringspaces=false,
    commentstyle=\color{olive},
    keywordstyle=\color{blue},
    identifierstyle=\color{black},
    stringstyle=\color{black},
    basicstyle=\ttfamily\footnotesize,
    breakatwhitespace=false,         
    breaklines=true,                 
    captionpos=b,                    
    keepspaces=true,    
    showspaces=false,                
    showstringspaces=false,
    showtabs=false,                  
    tabsize=2,
    emph={[2]optimal_cut, execute_optimal_cut, gate_cutting},
    emphstyle={[2]\color{purple}}
}
\begin{document}

\title{A Semantic Quantum Circuit Cache for Scalable and Distributed Quantum-Classical Workflows}



\author{
    \IEEEauthorblockN{Mar Tejedor\IEEEauthorrefmark{1}\IEEEauthorrefmark{2},  Javier Conejero\IEEEauthorrefmark{1}, Rosa M. Badia\IEEEauthorrefmark{1}\IEEEauthorrefmark{2}}\\
    \IEEEauthorrefmark{1}Barcelona Supercomputing Center, Barcelona, Spain\\
    \IEEEauthorrefmark{2}Universitat Politècnica de Catalunya, Barcelona, Spain\\
    mar.tejedor@bsc.es\\
}

\maketitle

\begin{abstract}
Hybrid quantum--classical workflows often execute large ensembles of circuits that differ syntactically but implement identical operations, leading to substantial redundant computation. To address this, we introduce the \emph{Quantum Circuit Cache}, a content-addressable system that detects semantic equivalence and reuses previously computed results across executions, backends, and workflow stages.  

Our approach combines ZX-calculus reduction with isomorphism-invariant Weisfeiler--Leman graph hashing to generate deterministic circuit identifiers, enabling constant-time lookup (O(1)) in distributed caches supporting both lightweight LMDB and scalable Redis deployments. The system integrates transparently into hybrid HPC workflows and remains backend-agnostic across CPU, GPU, and QPU environments.  

We evaluate the system on MareNostrum~5 with two representative workloads: distributed wire cutting and Differential Evolution–based QAOA optimization. For wire cutting, caching eliminates up to $91.98\%$ of redundant subcircuit simulations (equivalent to $7{,}544$ subcircuits), yielding speedups up to $7.0\times$ on a single node and maintaining advantages at scale, with Redis-based caching achieving up to $1.6\times$ speedups under high parallelism. Validation on a 35-qubit superconducting QPU (MareNostrum~Ona) confirms these benefits, achieving an $11.2\times$ speedup on real hardware. In distributed QAOA optimization, equivalence-aware caching avoids up to $27.6\%$ of circuit evaluations ($7{,}044$ circuits) and consistently reduces execution cost without altering the optimization algorithm. In both cases, reuse grows with concurrency and circuit structure, highlighting redundancy as a major systems bottleneck and demonstrating the effectiveness of our Quantum Circuit Cache.

These results establish semantic circuit caching as a powerful systems-level optimization for large-scale hybrid quantum--classical computing. As quantum computing increasingly integrates with HPC infrastructures, such techniques will play a central role in enabling efficient, scalable, and practical quantum applications on current and future platforms.
\end{abstract}

\input{Sections/1_Introduction}  
\input{Sections/2_State_of_art}
\input{Sections/3_Background}
\input{Sections/4_Quantum_Circuit_Cache}
\input{Sections/5_Evaluation}
\input{Sections/5.1_Evaluation_WC}

\input{Sections/5.2_Evaluation_QAOA}
\input{Sections/6_Discussion}
\input{Sections/7_Conclusions}

\section*{Acknowledgments}
The authors acknowledge funding from the Spanish Ministry for Digital Transformation and of Civil Service of the Spanish Government through the QUANTUM ENIA project call - Quantum Spain, EU through the Recovery, Transformation and Resilience Plan – NextGenerationEU within the framework of the Digital Spain 2026. The work has also been supported by the projects CEX2021-001148-S, and PID2023-147979NB-C21 from the MCIN/AEI and MICIU/AEI /10.13039/501100011033 and by FEDER, UE, by the Departament de Recerca i Universitats de la Generalitat de Catalunya, research group MPiEDist (2021 SGR 00412). In addition, the authors received financial support from the European Union, through the EuroHPC Joint Undertaking and its members under the Grant Agreement Nº 101159808, including top-up funding by the Ministry for Digital Transformation and the Civil Service of the Spanish Government. For M.T., the project that gave rise to these results received the support of a fellowship from the ”la Caixa” Foundation (ID 100010434). The fellowship code is LCF/BQ/DFR25/12000207.









\bibliographystyle{unsrt}
\bibliography{bibliography}

\end{document}

%% file: Sections/1_Introduction.tex
\section{Introduction}
\label{sec:introduction}
Hybrid quantum--classical computing has emerged as a dominant execution paradigm for near-term quantum algorithms, combining quantum processors with classical simulation, optimization, and orchestration layers.  
In practice, such workflows are increasingly executed at scale on High-Performance Computing (HPC) systems, where a significant number of quantum circuits may be generated, transformed, and executed as part of a single application.

A key, but often overlooked inefficiency in these workflows, is the prevalence of redundant quantum circuit execution.  
Redundancy arises naturally in a wide range of settings, including circuit cutting, variational quantum algorithms, parameter sweeps, error mitigation protocols, and benchmarking studies.  
In these scenarios, many circuits differ syntactically (e.g., due to gate reordering, compilation artifacts, or local simplifications) while remaining semantically equivalent -- they implement the same quantum operation.
Naïvely re-executing such circuits, whether via classical simulation on CPUs and GPUs or via execution on quantum processing units (QPUs), incurs significant computational cost, possible queueing delays, and unnecessary consumption of scarce quantum and classical resources.

Despite this, most existing quantum software stacks treat circuits as simple objects, identified only by their syntactic representation.  
This limitation becomes particularly acute at scale, where hybrid workflows generate massive numbers of overlapping circuits and where even modest reductions in redundant execution can yield substantial performance gains.

To address this challenge, we introduce the \emph{Quantum Circuit Cache}, a content-addressable caching subsystem designed to eliminate redundant quantum circuit execution through semantic equivalence detection.  
Rather than relying on syntactic equality, the cache identifies when a newly submitted circuit represents a quantum operation that has already been executed, and transparently retrieves the corresponding stored result.  
When a cache hit occurs, circuit execution (including translation, simulation, or hardware submission) is bypassed entirely.

The Quantum Circuit Cache is founded on three core principles.  
First, \emph{semantic equivalence}: equivalence detection is based on the underlying quantum operation, not on surface-level circuit syntax.  
Second, \emph{determinism and reproducibility}: cache keys are uniquely determined across runs, nodes, and execution environments, enabling reliable reuse in distributed settings.  
Third, \emph{scalability and backend independence}: the cache is designed to operate efficiently under HPC parallelism and to support heterogeneous simulation and execution backends, including CPUs, GPUs, and QPUs.

At the technical level, the cache combines graph-theoretic reduction using the ZX-calculus with isomorphism-invariant graph hashing to construct deterministic identifiers for quantum circuits.  
These identifiers are used to index a persistent key--value store, enabling constant-time lookup O(1) and retrieval of previously computed results.  
The system supports both lightweight, disk-backed deployments for local or moderate-scale workloads and distributed, in-memory deployments for large-scale HPC environments, while maintaining a unified cache format across backends.

In summary, this work makes the following contributions:
\begin{itemize}
    \item We identify redundant quantum circuit execution as an inefficiency in large-scale hybrid quantum--classical workflows.
    \item We propose a content-addressable Quantum Circuit Cache that enables semantic equivalence detection and transparent reuse of quantum circuit results.
    \item We present a graph-based reduction and hashing pipeline that yields deterministic, scalable circuit identifiers suitable for HPC deployment.
    \item We demonstrate a flexible cache architecture supporting both local and distributed backends, with portable persistence across execution environments.
\end{itemize}

Beyond a single application domain, this work positions semantic circuit caching as a reusable systems primitive for hybrid quantum--classical computing. By identifying computational equivalence at the circuit level rather than at the syntactic level, the proposed approach enables transparent reuse across optimization loops, distributed executions, and heterogeneous backends without modifying quantum algorithms or optimizers. To our knowledge, this is the first explicit implementation.

The remainder of this paper is organized as follows. Section~\ref{sec:state_of_art} reviews related work on quantum circuit equivalence and existing caching approaches. Section~\ref{sec:motivation_background} provides necessary background on quantum circuits, the ZX-calculus, and graph hashing techniques. Section~\ref{sec:quantum_circuit_cache} presents the design of our Quantum Circuit Cache, including the semantic hashing pipeline and distributed backend architecture. Section~\ref{sec:evaluation} evaluates the system on two representative hybrid quantum-classical workloads: distributed wire cutting (Section~\ref{subsec:wire_cutting_cache}) and Differential Evolution-based QAOA optimization (Section~\ref{subsec:qaoa_eval}). Section~\ref{sec:discussion} discusses the implications of our results and potential limitations. Finally, Section~\ref{sec:conclusion} concludes the paper and outlines directions for future work.

%% file: Sections/2_State_of_art.tex
\section{Related Work}
\label{sec:state_of_art}

Quantum circuit equivalence has been extensively studied in compilation, verification, and optimization, typically as a \emph{decision problem} aimed at validating compiler transformations or reasoning formally about circuit behavior. Early approaches relied on direct matrix comparison or symbolic reasoning, which rapidly becomes infeasible for large circuits \cite{doi:10.1142/S0219749905001067}. To improve scalability, structural representations and classical reasoning techniques were introduced. Decision-diagram methods such as quantum multiple-valued decision diagrams (QMDDs) provide compact operator representations for equivalence checking without full matrix construction, though memory can grow rapidly for deep or highly entangled circuits~\cite{li2022quantummultiplevalueddecisiondiagrams, 7163590}. More advanced techniques combine symbolic reasoning with canonical representations to improve robustness in compiler validation~\cite{Burgholzer2020Advanced}, and SAT- or model-counting-based approaches achieve strong performance in restricted regimes but remain costly for general circuits \cite{10.1145/3729229}.

Graphical reasoning using the ZX-calculus has emerged as a powerful framework for circuit equivalence and simplification. The ZX-calculus represents quantum processes diagrammatically with semantics-preserving rewrite rules that reveal nontrivial circuit identities~\cite{backens2014zx}. Automated tools such as \texttt{PyZX} \cite{kissinger2020pyzx} enable substantial reductions through algebraic simplification and normalization~\cite{duncan2020graphical}, and ZX-based techniques have been used to verify compiler flows and detect equivalence in variational algorithms~\cite{Peham2022Equivalence}. While effective for optimization and correctness verification, these methods are not designed for persistent reuse across executions.  

Our approach differs by promoting equivalence detection from a verification tool to a systems primitive for hybrid quantum--classical workflows. By combining ZX-calculus reduction with isomorphism-invariant graph hashing, we generate deterministic identifiers tied to computational semantics rather than syntactic structure. These identifiers support constant-time lookup within a persistent circuit cache, enabling reuse across executions, optimization iterations, circuit cutting workflows, and heterogeneous backends. To our knowledge, this is the first explicit implementation of a quantum circuit cache designed to reuse previously executed quantum computations at scale.

%% file: Sections/3_Background.tex

\begin{figure*}[t!]
    \centering
    \includegraphics[width=0.9\linewidth]{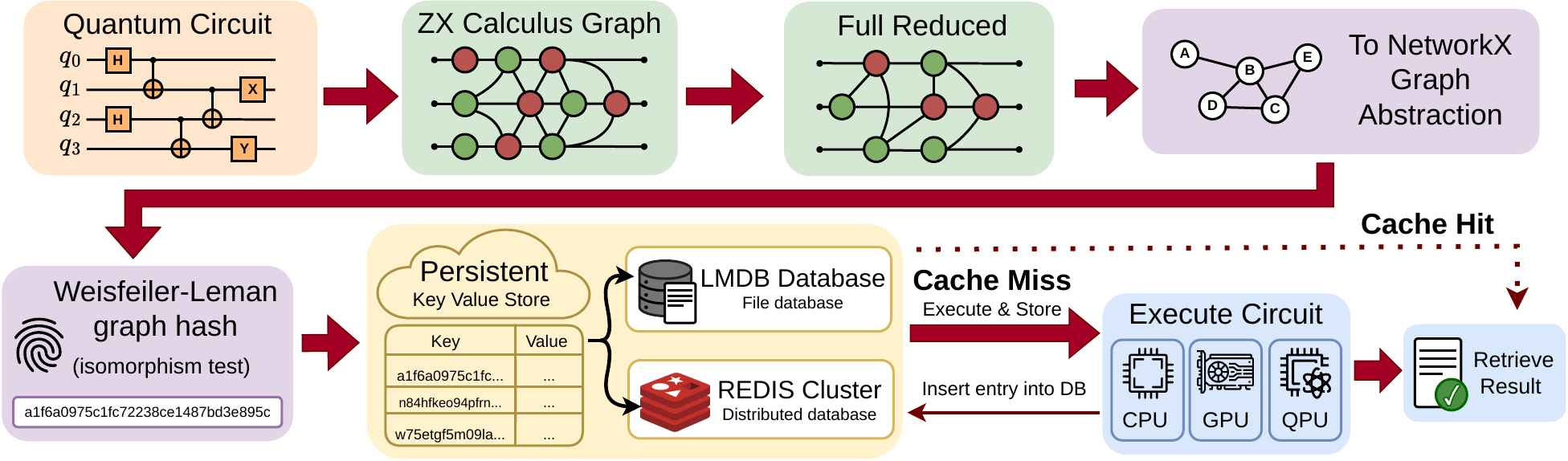}
    \caption{End-to-end workflow of the Quantum Circuit Cache. Incoming circuits are translated into ZX-calculus graphs, reduced via a deterministic rewrite strategy, deterministically serialized, and hashed using Weisfeiler--Leman refinement. Cache hits bypass execution and retrieve stored results, while cache misses trigger execution on CPU, GPU, or QPU backends and insertion into a persistent key--value store.}
    \label{fig:quantum_circuit_cache_workflow}
\end{figure*}


\section{Motivation and Background}
\label{sec:motivation_background}

Current quantum software lacks a robust notion of semantic circuit identity. Circuits are typically represented as gate sequences, QASM strings, or compiler-specific data structures that are sensitive to ordering and optimizations, so semantically equivalent circuits are often treated as distinct. Exact equivalence checking via matrix comparison or symbolic reasoning is computationally infeasible in runtime pipelines, while identifiers must remain deterministic and reproducible across distributed nodes to enable safe reuse. In hybrid HPC-quantum scale environments, where thousands of circuit evaluations run concurrently on CPUs, GPUs, and quantum processing units, any practical reuse mechanism must provide deterministic identifiers, constant-time lookup, concurrency-safe access, and persistence across runs and environments.

A practical solution requires a representation that captures circuit semantics while being robust to syntactic variation. Graphical representations based on the ZX-calculus provide such a foundation, enabling semantics-preserving rewrites that eliminate superficial differences without exhaustive equivalence proofs. After reduction, circuits are assigned stable identifiers via isomorphism-invariant graph hashing. We employ Weisfeiler--Leman algorithm \cite{WL} to generate compact, deterministic fingerprints that enable distributed content-addressable storage.

Detecting reusable quantum computations is fundamentally more challenging than classical memoization. Quantum circuits admit multiple syntactic realizations of the same unitary due to commutation relations, gate identities, phase cancellations, alternative decompositions, and compiler nondeterminism. Traditional hashing or compiler memoization fails because syntactic equality rarely holds, and existing transpilers do not produce backend-independent unique representations. Our approach therefore extends beyond caching, establishing a semantic identity layer that assigns identifiers based on computational meaning rather than compilation artifacts.

The proposed semantic hashing pipeline combines ZX-calculus simplification with Weisfeiler--Leman graph hashing \cite{WL}, balancing robustness and scalability. Equivalence detection is intentionally incomplete: ZX rewriting can prove many circuit identities, but exhaustive unitary checking is prohibitive. Semantically equivalent circuits may occasionally produce different hashes if reductions converge to different forms, which reduces reuse opportunities but never compromises correctness. The approach is particularly effective for circuit families dominated by Clifford+T structures, commuting rotations, layered variational ansätze, and circuit-cutting fragments, where ZX reductions converge consistently. By prioritizing scalability over formal completeness, the pipeline provides a reliable but conservative mechanism that maximizes reuse while preserving correctness.

Taken together, these considerations motivate treating semantic equivalence as a systems primitive rather than a verification task. Instead of explicitly proving equivalence, the goal is to construct deterministic semantic identifiers that enable efficient and fast, transparent reuse of previously executed quantum computations under massive parallelism. The Quantum Circuit Cache realizes this concept through semantics-preserving, deterministic normalization and HPC-compatible persistent storage, enabling reuse across workflow stages, execution backends, and repeated executions.

%% file: Sections/4_Quantum_Circuit_Cache.tex

\begin{table*}[t]
    \centering
    \caption{Comparison of database backends used in the Quantum Circuit Cache.}
    \resizebox{0.9\textwidth}{!}{
    \begin{tabular}{lcc}
        \toprule
        Feature & LMDB & Redis Cluster \\
        \midrule
        Target scale & Local / small HPC & Large HPC / distributed \\
        Concurrent readers & Multiple & Multiple \\
        Concurrent writers & Single (persistent task) & Multiple \\
        Memory footprint & Low & Medium--High \\
        Persistence & On-disk & In-memory (optional LMDB export) \\
        Deployment complexity & Low & High \\
        Best suited for & Lightweight, deterministic workflows & High-parallelism workloads \\
        \bottomrule
    \end{tabular}
    }
    
    \label{tab:cache_db_comparison}
\end{table*}


\section{Quantum Circuit Cache}
\label{sec:quantum_circuit_cache}

In order to address unnecessary computational cost, possible queueing delays, and resource consumption inefficiency, we introduce the \emph{Quantum Circuit Cache}, a content-addressable caching subsystem that enables semantic equivalence detection and transparent reuse of quantum circuit results.

Figure~\ref{fig:quantum_circuit_cache_workflow} illustrates the end-to-end cache workflow.  
Direct comparison of circuits at the gate-list or QASM level is insufficient for equivalence detection, as such representations are highly sensitive to syntactic variation.
Instead, circuits are deterministically translated into ZX-calculus graphs using the \texttt{PyZX} library, yielding graphs composed of spider vertices, edges, and phase annotations that fully capture the implemented quantum operation.
Semantically equivalent circuits may still produce distinct ZX graphs; therefore, each graph undergoes a \emph{Full Reduce} procedure from the same \texttt{PyZX} library, consisting of a sequence of semantics-preserving rewrite rules such as spider fusion, identity elimination, phase normalization, local complementation, and pivoting \cite{vandeWetering:2020giq}

Although the ZX-calculus does not guarantee a unique normal form, \textit{Full Reduce} substantially reduces structural variability and collapses many syntactically distinct but semantically equivalent circuits into a common reduced representation.
In practice, this eliminates differences arising from gate reordering, local simplifications, and compiler-induced transformations.  
The computational cost of this reduction is amortized across repeated cache accesses: once a circuit has been reduced and cached, subsequent equivalent circuits incur only constant-time lookup overhead, which is especially beneficial in workloads where identical subcircuits recur frequently.

Following reduction, the ZX graph is converted into a deterministic, backend-independent representation. Vertex attributes (type and phase) and adjacency information are encoded using a fixed ordering and stable attribute encoding.
To decouple the cache from ZX-specific data structures, this form is a \texttt{NetworkX}~\cite{NetworkX} graph, which serves as a uniform abstraction layer between quantum-specific representations and classical graph representations.

From this graph representation, a deterministic identifier is constructed using the Weisfeiler--Leman (WL) graph hashing function provided by the \texttt{NetworkX} library.
The idea behind WL is simple: it repeatedly summarizes each node by aggregating information from its neighbors, progressively refining these local summaries into more expressive structural labels.  
After several iterations, this process produces a compact fingerprint (sixteen character string) of the entire graph that captures its structure and is invariant under graph isomorphism.
We use this implementation directly to generate the cache key.  
Although WL hashing is not guaranteed to be collision-free, in practice collisions are extremely rare for the reduced ZX graphs arising in realistic quantum workloads. In the unlikely event of a collision, we recommend storing metadata alongside each cache entry (e.g., circuit depth or number of qubits) that can be used to validate whether a retrieved result corresponds to the submitted circuit, gracefully falling back to execution if a mismatch is detected.
Compared to exact graph isomorphism testing, WL hashing is significantly cheaper computationally, reducing the problem to near constant-time lookup and making large-scale reuse practical.

The Quantum Circuit Cache operates as a content-addressable key--value store indexed by these WL hashes.  
Each cache entry may store simulation results, measurement statistics, expectation values, or execution metadata such as backend type, number of shots, and noise model parameters.
Cache keys are backend-agnostic: a single circuit hash may be associated with multiple backend-specific results, enabling reuse across heterogeneous execution targets and supporting hybrid quantum--classical workflows.

To support different deployment settings, we provide two complementary cache backends.  
The first uses the \emph{Lightning Memory-Mapped Database} (LMDB)~\cite{LMDB}, a fast key--value store designed for workloads dominated by reads.  
LMDB maps the database directly into memory, which enables fast access and reduces runtime overhead, making it suitable for local execution and small-to-medium HPC deployments.
However, because LMDB only allows a single writer at a time, we introduce a dedicated \emph{persistent writer task} to ensure safe concurrent access.
In this design, parallel tasks write new cache entries atomically to an intermediate file that acts as a queue, relying on filesystem atomic write guarantees, while the persistent task continuously consumes these entries and updates the database.

For large-scale distributed workloads, an alternative backend based on a Redis cluster~\cite{Web:redis, Web:redis_cluster} is provided.  
Redis supports multiple concurrent readers and writers, automatic sharding, replication, and high-throughput in-memory access, enabling scalable cache access across many nodes without centralized coordination.
Each compute node connects directly to the Redis cluster, with concurrency and data distribution handled internally, at the cost of increased memory usage and deployment complexity.

To summarize the trade-offs between both backends, Table~\ref{tab:cache_db_comparison} compares their key characteristics.  
While LMDB provides a lightweight and deterministic solution for moderate parallelism, Redis enables high-concurrency distributed access at scale.
Both backends share identical cache semantics and can interoperate via a common persistence mechanism.

To enable portability and long-term reuse, we implement a cross-backend persistence mechanism using LMDB as a universal exchange format.  
At the end of a workflow, the contents of a Redis cluster can be exported into an LMDB database, which can then be used to initialize future executions regardless of the chosen backend.
Unlike Redis native persistence mechanisms, which require restoring an identical cluster topology, the LMDB format is self-contained and backend-agnostic.
Although exporting and reloading the cache incurs overhead proportional to its size, this cost is amortized over repeated executions and is outweighed by the benefits in reproducibility, portability, and deployment flexibility.

Overall, the Quantum Circuit Cache provides a semantics-aware, scalable, and backend-independent mechanism for eliminating redundant quantum circuit execution.  
By combining ZX-calculus-based canonicalization, isomorphism-invariant graph hashing, and HPC-compatible storage backends, the cache integrates seamlessly into hybrid quantum--classical workflows and delivers significant performance gains across a wide range of execution environments. 

%% file: Sections/5_Evaluation.tex
\section{Evaluation}
\label{sec:evaluation}

This section evaluates the Quantum Circuit Cache on two representative hybrid quantum-classical workloads: distributed wire cutting and Differential Evolution-based QAOA optimization. Both workloads are executed on MareNostrum 5, a large-scale HPC system located at the Barcelona Supercomputing Center (BSC). Each compute node provides 112 CPU cores (Intel Xeon Platinum 8360Y, 2.4 GHz) and 256 GB of main memory. For QPU validation, we use MareNostrum Ona, a 35-qubit superconducting quantum processor, also located at BSC. All experiments use the PyCOMPSs task-based runtime (version 3.3.3) to orchestrate task scheduling and data dependencies across nodes, enabling high-throughput parallel execution of quantum circuit simulations. Circuit simulations are performed using the Qiskit Aer statevector simulator (version 0.14.2) with Qiskit (version 1.1.1), which provides exact statevector evolution suitable for evaluating caching correctness and performance without confounding noise effects.

The two workloads are chosen to stress different sources of semantic redundancy. Wire cutting generates a combinatorial explosion of subcircuits that differ only in classical coefficients while the underlying quantum operations most remain identical, making it an ideal stress test for structural reuse. In contrast, QAOA with Differential Evolution explores a continuous parameter space, where semantic equivalence arises from parameter discretization and ZX-calculus simplifications that collapse distinct parameter vectors into identical canonical forms. Together, these workloads demonstrate that semantic circuit caching is effective across qualitatively different redundancy patterns.

All experiments are fully reproducible and the complete source code will be available here\footnote{\url{https://github.com/bsc-wdc/qdislib}} in the next Qdislib release. The following software versions are used throughout: Qiskit 1.1.1, Qibochem 0.0.4, PyZX 0.9.0, NetworkX 3.3, LMDB 1.7.3, and redis-py 5.0.3. All random number generators are seeded for reproducibility as detailed below.

%% file: Sections/5.1_Evaluation_WC.tex
\subsection{Wire Cutting with Distributed Circuit Caching}
\label{subsec:wire_cutting_cache}

Wire cutting decomposes a global quantum circuit into multiple subcircuits by cutting selected wires, at the cost of decomposing the corresponding quantum channel. Formally, each cut wire is expanded into a complete operator basis, expressed as a linear combination of measurement-preparation operations. In the single-qubit case, this basis is given by the Pauli operators $\{I, X, Y, Z\}$, each with two eigenvalues \cite{PhysRevLett.125.150504}. This produces up to 8 combinations with a total of 16 distinct independent subcircuits for recreating one cut.

When multiple cuts are introduced, all combinations of these basis elements must be considered. For four cuts, this yields $8^4 = 4096$ combinations, which doubles to 8192 subcircuits because each cut splits the original circuit into two fragments. Each term corresponds to a distinct subcircuit instance defined by a specific choice of measurement and preparation operators applied at the cut locations. The expectation value of the original circuit is then reconstructed by classically combining the results of all subcircuits, weighted by the corresponding decomposition coefficients.

This expansion produces a combinatorial number of reconstruction terms that often differ only in classical coefficients, while the underlying quantum subcircuits remain identical. Consequently, wire cutting inherently generates large amounts of redundant computation, an effect that grows with the number of cuts and the degree of parallelism.

Distributed circuit caching addresses this redundancy by storing previously simulated subcircuits and reusing them whenever an identical subcircuit is encountered. In this evaluation, we store the full statevector (complex128 array of length $2^{n}$, where $n$ is the number of qubits) for each cached subcircuit, as wire cutting requires full probability distributions or expectation values computed from statevectors. This avoids re-executing identical quantum workloads and substantially reduces overall resource consumption.

\begin{figure}[b]
    \centering
    \includegraphics[width=\linewidth]{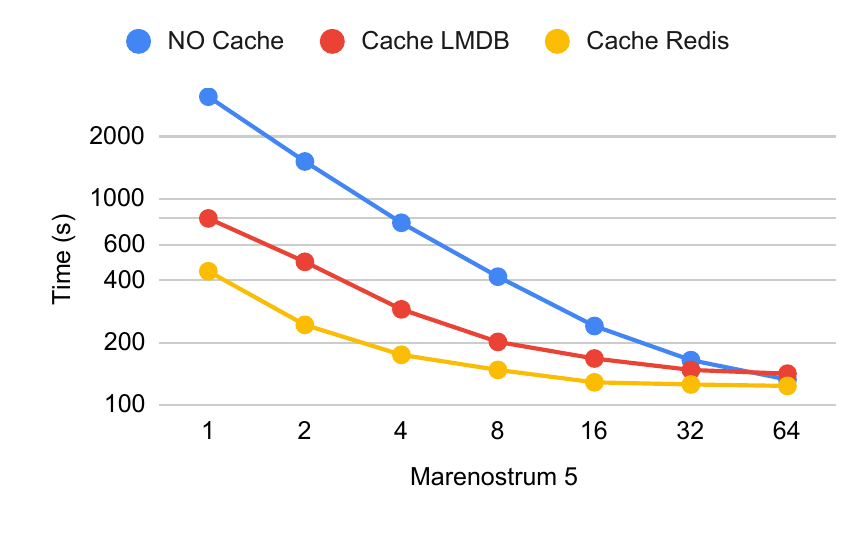}
    \caption{Total execution time for HEA circuits with four wire cuts as a function of compute nodes. Distributed circuit caching substantially reduces runtime, with Redis consistently outperforming LMDB at larger scales.}
    \label{fig:hea_runtime}
\end{figure}

To evaluate the cache under increasing parallel pressure, we scale the execution from 1 to 64 compute nodes, doubling at each step. All subcircuits are simulated exactly using the Qiskit Aer statevector simulator, returning the full statevector as the cached result. We consider two representative circuit families, each with 48 qubits and four wire cuts per circuit. Due to the wire cutting procedure, each cut increases the effective circuit size by introducing ancilla qubits required to mediate the measurement-preparation decomposition, resulting in subcircuits that contain up to 28 qubits: Hardware-Efficient Ansatz (HEA) circuits with 2 layers of alternating rotation and entanglement gates generated by Qibochem ansatz \cite{qibochem}, and random circuits generated using Qiskit's \texttt{random\_circuit} function with \texttt{depth=4}, \texttt{max\_operands=2}, \texttt{measure=False}, \texttt{conditional=False}, and \texttt{seed=1000}, where each random gate is assigned a parameter value uniformly sampled from $[0, 2\pi)$ using a seed.

\begin{figure}[htbp]
    \centering
    \begin{subfigure}{0.48\linewidth}
        \centering
        \includegraphics[width=\linewidth]{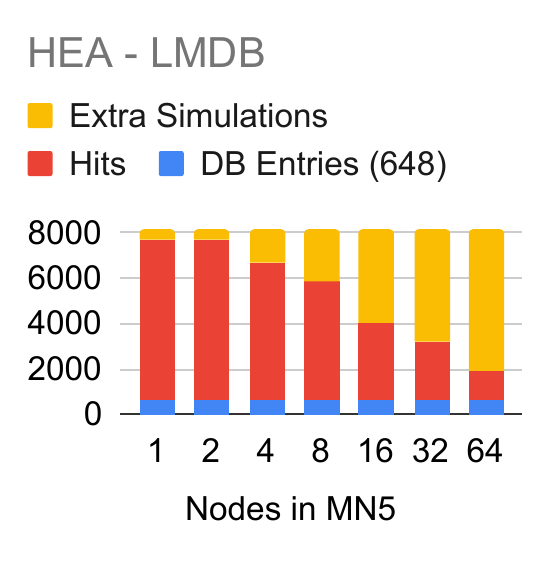}
        \caption{Cache behavior for HEA circuits using an LMDB backend. Cache hits dominate, but extra simulations increase with parallelism due to the single-writer constraint.}
        \label{fig:hea_cache_lmdb}
    \end{subfigure}
    \hfill
    \begin{subfigure}{0.48\linewidth}
        \centering
        \includegraphics[width=\linewidth]{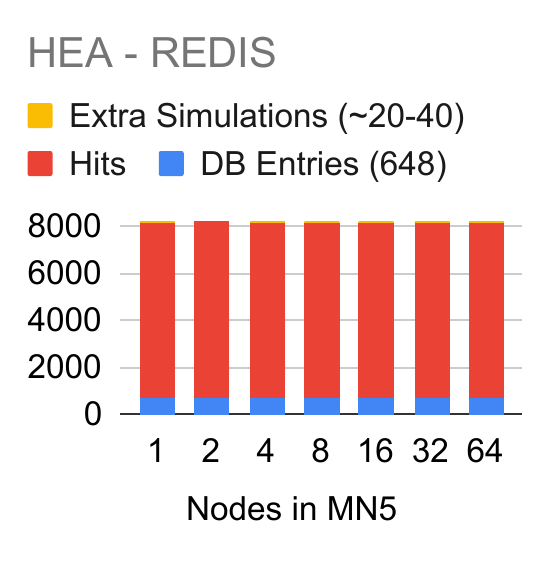}
        \caption{Cache behavior for HEA circuits using a Redis backend. Concurrent writes significantly reduce extra simulations, enabling stable scaling under high parallelism.}
        \label{fig:hea_cache_redis}
    \end{subfigure}

    \caption{Comparison of cache behavior for HEA circuits using LMDB and Redis backends. Y axis is the total number of sucircuits simulations done.}
    \label{fig:hea_cache_comparison}
\end{figure}

We compare three execution configurations: execution without caching (baseline), caching using an LMDB backend with a centralized write queue, and caching using a Redis-based distributed cluster. The same random seeds are used across caching and no-caching runs to ensure fair comparison.

Figures~\ref{fig:hea_runtime}--\ref{fig:hea_cache_redis} summarize results for HEA circuits. Figure~\ref{fig:hea_runtime} shows total execution time as a function of compute nodes. Without caching, runtime decreases with scale due to parallel execution, but enabling caching produces substantial additional speedups at all scales. On a single node, Redis achieves a $7.0\times$ speedup and LMDB achieves $3.9\times$. At four nodes, Redis has a $4.4\times$ speedup, while LMDB reaches $2.6\times$. At sixteen nodes, Redis delivers a $1.9\times$ speedup compared to $1.4\times$ for LMDB. At thirty-two nodes, Redis continues to provide a measurable benefit with a $1.3\times$ speedup, while LMDB converges to $1.1\times$.

Figures~\ref{fig:hea_cache_lmdb} and~\ref{fig:hea_cache_redis} break down cache behavior. Each bar decomposes subcircuit executions into cache hits, database entries, and extra simulations caused by concurrent insertion attempts. For HEA circuits, cache hits dominate across all scales, achieving a hit rate of $91.98\%$, theoretically avoiding $7{,}544$ subcircuit simulations (total $8{,}192$ possible subcircuits minus $648$ unique entries). The LMDB backend shows an increasing fraction of extra simulations as parallelism grows due to its single-writer constraint, limiting scalability at high node counts. In contrast, Redis supports concurrent writes and distributed coordination, suppressing redundant execution (only 20–40 extra simulations) and preserving speedup under high parallelism. The database ultimately stores $648$ unique circuit entries.

\begin{figure}[htbp]
    \centering
    \includegraphics[width=0.9\linewidth]{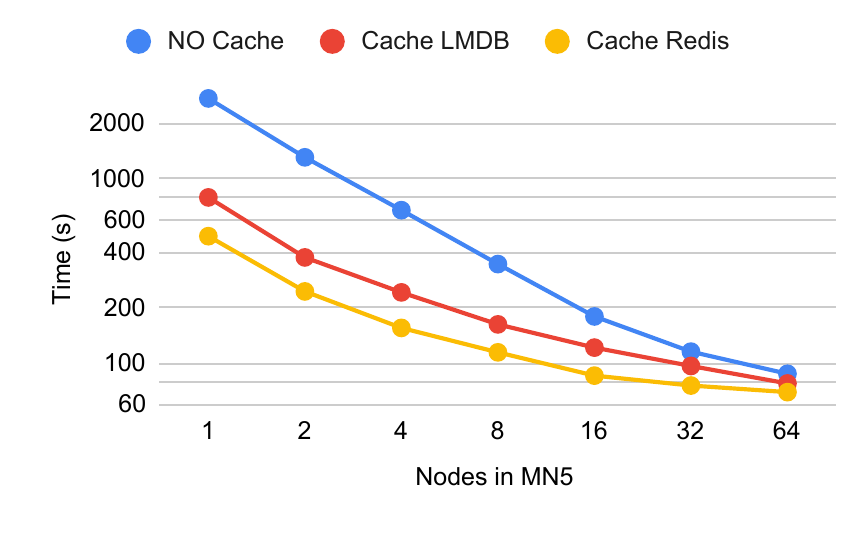}
    \caption{Total execution time for random circuits with four wire cuts as a function of compute nodes.}
    \label{fig:random_runtime}
\end{figure}

\begin{figure}[htbp]
    \centering
    \begin{subfigure}{0.48\linewidth}
        \centering
        \includegraphics[width=\linewidth]{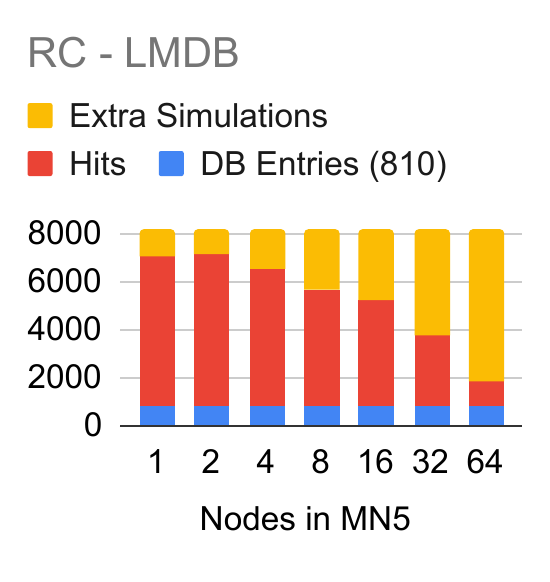}
        \caption{Cache behavior for random circuits using an LMDB backend. Cache hits are significant, but extra simulations increase with scale.}
        \label{fig:random_cache_lmdb}
    \end{subfigure}
    \hfill
    \begin{subfigure}{0.48\linewidth}
        \centering
        \includegraphics[width=\linewidth]{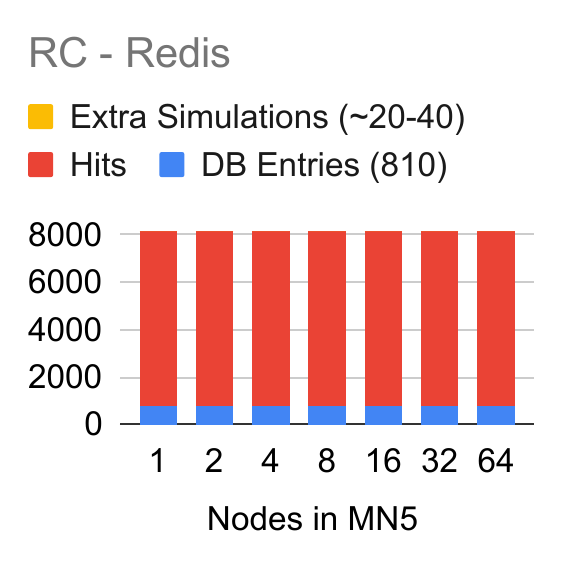}
        \caption{Cache behavior for random circuits using a Redis backend. Concurrent access reduces extra simulations and stabilizes performance.}
        \label{fig:random_cache_redis}
    \end{subfigure}

    \caption{Comparison of cache behavior for random circuits (RC) using LMDB and Redis backends. Y axis is the total number of sucircuits simulations done.}
    \label{fig:random_cache_comparison}
\end{figure}

Figures~\ref{fig:random_runtime}--\ref{fig:random_cache_redis} report results for a single instance of a random circuit. Although random circuits lack the strong structural regularity of HEA circuits, caching still yields unexpectedly high speedups. This is a significant finding: structural regularity is not a prerequisite for effective caching. Even in highly unstructured random circuits, substantial reusable structure is captured, leading to consistently strong performance improvements.

As shown in Figure~\ref{fig:random_runtime}, on a single node, Redis achieves a $5.6\times$ speedup, while LMDB reaches $3.4\times$. At four nodes, Redis maintains a $4.3\times$ speedup compared to $2.8\times$ for LMDB. At sixteen nodes, Redis achieves a $2.1\times$ speedup versus $1.5\times$ for LMDB. At thirty-two nodes, Redis delivers a $1.5\times$ speedup compared to $1.2\times$ for LMDB. At the largest scale (64 nodes), both backends converge to a similar $1.3\times$ (Redis) and $1.1\times$ (LMDB).

\begin{table}[b]
\small
\centering
\caption{Average execution time (seconds) per pipeline stage.}
\begin{tabular}{lcc}
\toprule
Stage & Cache Miss  \\
\midrule
Qiskit $\rightarrow$ ZX conversion & 0.02 \\
ZX $\rightarrow$ NetworkX graph & 0.002 \\
WL hashing & 0.006 \\
Cache lookup & 0.064 \\
Simulation & 35.48 \\
Cache store & 0.043  \\
\midrule
Total overhead (no simulation) & $\approx 0.13$  \\
\bottomrule
\end{tabular}
\label{tab:pipeline_overhead}
\end{table}

\begin{table*}[t]
\small
\centering
\caption{Quantum hardware validation results on MareNostrum~Ona, a 35-qubit superconducting quantum processor.}
\label{tab:qpu_validation}
\begin{tabular}{lccccc}
\toprule
Configuration & Unique Circuits & Total Circuits & QPU Time (with cache) & QPU Time (without cache) & Speedup \\
\midrule
4 cuts (HEA, 2 layers) & 648 & 8192 & 1.83 hours & $\sim$20.5 hours$^*$ & 11.2$\times$ \\
2 cuts (HEA, 1 layer) & 36 & 128 & 5.86 min & 17.5 min & 2.98$\times$ \\
\bottomrule
\end{tabular}
\par\vspace{4pt}
\footnotesize $^*$Theoretical estimate based on simulation (8,192 circuits × 9s). The 2-cuts without cache was measured directly on QPU.
\end{table*}

Figure~\ref{fig:random_cache_comparison} breaks down the cache behavior. Cache hits remain significant across all scales, achieving a hit rate of $89.92\%$. The total number of possible subcircuits for four cuts is $8{,}192$, and the database stores $810$ unique circuit entries, theoretically avoiding $7{,}382$ subcircuit simulations ($8{,}192 - 810 = 7{,}382$). In practice, however, some redundant simulations still occur due to concurrent insertion attempts, particularly with the LMDB backend where the single-writer constraint causes extra simulations to increase with parallelism. Redis, by contrast, supports concurrent writes and distributed coordination, significantly reducing these redundant executions to only 20–40 extra simulations across hundreds of concurrent writers, stabilizing performance at higher node counts.

Table~\ref{tab:pipeline_overhead} presents the average execution time per pipeline stage. The full semantic identification pipeline (including Qiskit-to-ZX conversion, ZX-to-graph transformation, Weisfeiler--Leman hashing, cache lookup, and result storage) incurs an average overhead of approximately $0.13$ seconds per circuit on cache misses, independent of the database backend used.

Critically, this overhead is dominated by cache lookup latency, while the semantic reduction steps (ZX conversion and WL hashing) remain in the millisecond range. Circuit simulation of 28 qubits requires on average $35.48$ seconds, making the caching overhead roughly two orders of magnitude smaller than a single execution. This imbalance highlights the fundamental motivation of the system: even a modest cache hit rate amortizes the entire pipeline cost.

Furthermore, the breakdown shows that semantic reduction is highly efficient relative to storage operations, indicating that the proposed graph-based identity layer is not a bottleneck even under repeated large-scale execution. The cache is computationally lightweight enough to be deployed transparently in HPC workflows, where its overhead is negligible compared to both classical simulation and QPU times.

These results (combination of high hit rates, low overhead, and scalable backend support) makes the Quantum Circuit Cache highly effective for wire-cutting workloads across different circuit families. HEA circuits benefit most strongly due to their inherent structural, while random circuits also exhibit substantial performance improvements despite lacking explicit structure. By eliminating redundant execution of identical subcircuits ($7{,}544$ for HEA and $7{,}382$ for random circuits) the caching mechanism significantly reduces both classical computational cost and overall execution workload.

To validate the cache on actual quantum hardware, we executed wire cutting workloads on MareNostrum~Ona, a 35-qubit superconducting quantum processor at the Barcelona Supercomputing Center. The experimental setup mirrors our simulation-based evaluation: a classical HPC node handles circuit cutting and cache management, while the QPU executes the resulting subcircuits in series.

We evaluate two configurations, summarized in Table~\ref{tab:qpu_validation}. First, the 48-qubit HEA circuit with two layers and four cuts generates 8,192 subcircuits. Our semantic cache contained 648 unique subcircuits (matching the database size from our simulation). Executing all these 648 unique circuits on the QPU required 1.83 hours. This accounts only for QPU time, the end-to-end time would include the time overhead of circuit cutting, decomposition, and result reconstruction. However, this overhead is identical in both the cached and uncached scenarios.

Without caching, executing all 8,192 subcircuits would consume an estimated minimum of 20.5 hours of QPU time (8,192 × 9 seconds per circuit), a prohibitively expensive and wasteful use of limited quantum resources. We therefore report this as a theoretical estimate, as submitting 8,192 redundant circuits to the QPU would be both impractical and unnecessary given the cache's demonstrated effectiveness. The measured 1.83 hours with caching yields a $11.2\times$ speedup compared to this theoretical baseline.

Second, we tested a smaller HEA circuit (reduced to a single layer) with two cuts, generating 128 total subcircuits. The cache identified 36 unique circuits in the database with 8 extra simulations due to cache concurrency, which executed in 5.86 minutes on the QPU. In this case, we also measured the time without caching directly on the QPU: executing all 128 circuits required 17.5 minutes. This yields a $2.98\times$ speedup.

These results confirm that the Quantum Circuit Cache provides substantial benefits on real quantum hardware, where queue times and limited QPU availability make redundant execution particularly costly. The $11.2\times$ speedup for four cuts demonstrates that caching becomes increasingly valuable as circuit complexity and cut depth increase.

%% file: Sections/5.2_Evaluation_QAOA.tex
\subsection{Distributed Differential Evolution QAOA with Circuit Caching}
\label{subsec:qaoa_eval}

Variational quantum algorithms such as the Quantum Approximate Optimization Algorithm (QAOA) are naturally well suited to semantic caching, as they repeatedly evaluate parameterized quantum circuits inside a classical optimization loop. During optimization, identical or semantically equivalent circuits are often executed multiple times across iterations, across population members, and across different circuit depths. The objective of this evaluation is therefore not to improve algorithmic solution quality, but to quantify the systems-level impact and practical versatility of equivalence-aware circuit caching in a distributed variational workload.

We consider QAOA applied to the Max-Cut problem using a distributed Differential Evolution (DE) optimizer. DE is a population-based, gradient-free method that evaluates many candidate solutions per generation and frequently revisits similar regions of the parameter space due to its mutation and crossover operators. This high-throughput evaluation pattern makes it particularly suitable for stressing circuit reuse mechanisms, as many circuits are explored simultaneously and often revisit similar parameter regions. Prior work has demonstrated the effectiveness of evolutionary methods for parameterized quantum circuits \cite{farhi2014qaoa, ostaszewski2019evqe, failde2023de_vqa}.

From a systems perspective, population-based optimizers are especially relevant because each population member can be evaluated independently, producing large numbers of concurrent circuit evaluation requests and naturally stressing the caching subsystem.

We evaluate Max-Cut QAOA circuits on a random 24-vertex graph with 60 edges (generated with seed 42). QAOA depths are $p \in \{2,3,4\}$, and parameters are discretized onto fixed $(\beta,\gamma)$ grids:
\[
\beta \in \texttt{linspace}(0,\pi/2,N_\beta), \quad
\gamma \in \texttt{linspace}(0,2\pi,N_\gamma),
\]
using three discretization resolutions: coarse $(N_\beta=16, N_\gamma=32)$, medium $(32, 64)$, and fine $(64, 128)$. Discretization intentionally increases the probability that distinct parameter vectors map to identical circuit instances after ZX-calculus simplification.

The DE optimizer uses the \texttt{best1bin} strategy with population size 500, 50 generations, mutation factor $F=0.7$, crossover probability $CR=0.7$, and seed 100. We allocate one compute node per 100 population members, resulting in 5 nodes total. Each population member corresponds to a single QAOA circuit simulation task, and all tasks share a distributed circuit cache. Within each generation, all circuit evaluations execute in parallel.

Unlike the wire-cutting evaluation, we do not distinguish between LMDB and Redis backends here, as both exhibit identical cache-hit behavior under this workload. The node count is relatively small (5 nodes), and concurrency occurs only at the population level within each iteration, which does not introduce significant contention. Consequently, the extra simulations observed with LMDB under high parallelism are not present in this scenario.

\begin{figure}[htbp]
    \centering
    \includegraphics[width=\linewidth]{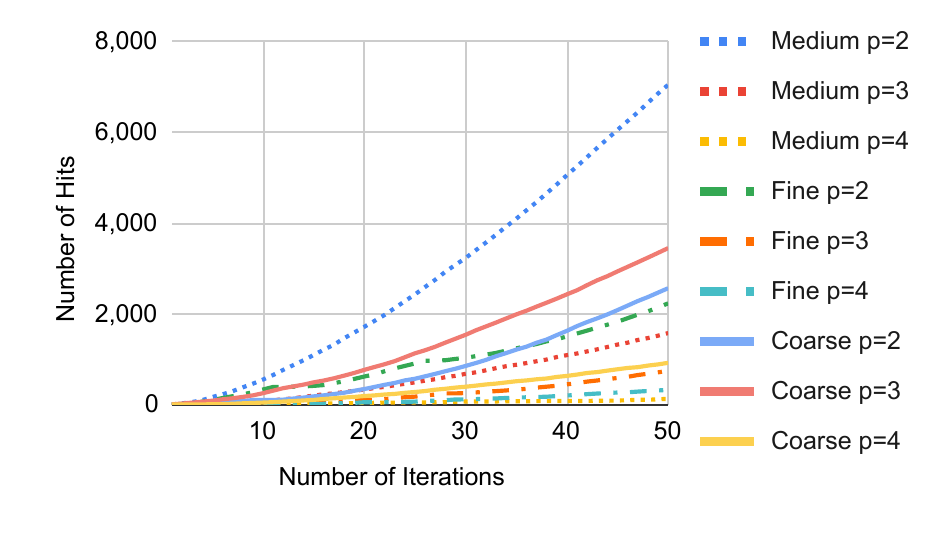}
    \caption{Cumulative cache hits versus DE iteration for QAOA depths $p=2,3,4$ under coarse, medium, and fine parameter discretization. Cache reuse increases steadily over time, with coarser and medium discretizations exhibiting higher hit rates.}
    \label{fig:qaoa_hits_iterations}
\end{figure}

Figure~\ref{fig:qaoa_hits_iterations} shows the cumulative number of cache hits as a function of DE iteration for different circuit depths and discretization resolutions. Across all configurations, cache hits increase monotonically as optimization progresses, indicating that the optimizer repeatedly revisits previously explored regions of the discretized parameter space. Coarser and medium discretizations produce faster growth in cache reuse, while deeper circuits exhibit delayed but still substantial reuse. For $p=2$ with medium discretization, the hit rate reaches $27.6\%$, the highest observed in this evaluation.

\begin{figure}
    \centering
    \includegraphics[width=\linewidth]{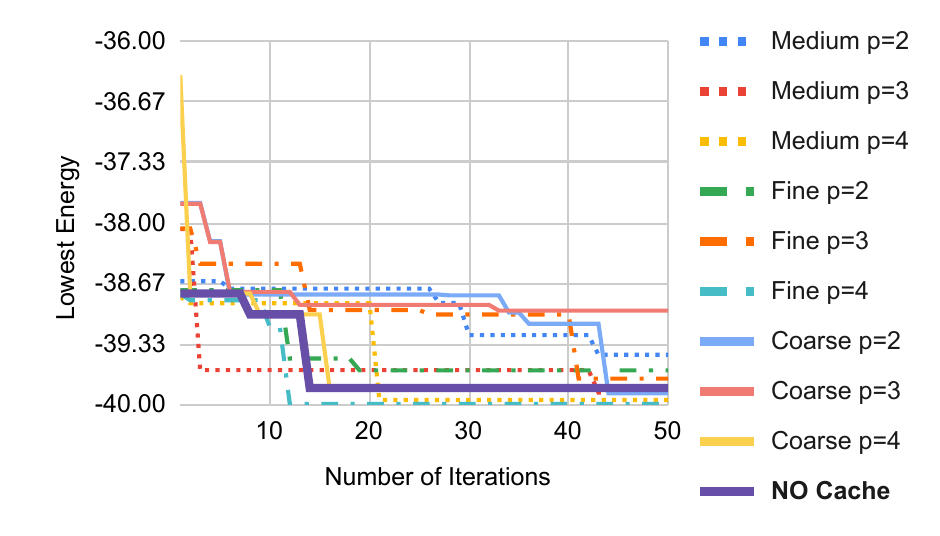}
    \caption{Best Max-Cut energy versus DE iteration for different QAOA depths and discretization resolutions. Energies converge rapidly and saturate within a narrow range due to equivalence-aware circuit evaluation.}
    \label{fig:qaoa_energy_iterations}
\end{figure}

Figure~\ref{fig:qaoa_energy_iterations} reports the best observed Max-Cut energy as a function of DE iteration. All runs converge rapidly and stabilize within a narrow energy range. While increasing circuit depth or refining discretization would in principle expand the search space, we do not observe consistent improvement in final energies across configurations. This behavior is partly explained by circuit equivalences: many parameter settings reduce to identical canonical forms after ZX simplification, effectively compressing the optimization landscape into equivalence classes. To verify that caching does not alter the optimization trajectory, we compared a cached run against a baseline without caching using the middle configuration ($p=4$, medium discretization). This baseline run produced nearly identical convergence behavior to the cached runs. Importantly, convergence trends remain stable across all configurations, indicating that caching eliminates redundant evaluations without adversely affecting optimizer behavior.

Aggregate cache statistics after the full optimization run are summarized in Table~\ref{tab:qaoa_stats}. Even under fine discretization, thousands of circuit evaluations are avoided, demonstrating that semantic redundancy persists even in relatively high-resolution parameter spaces. However, the relationship between discretization resolution and cache efficiency is not monotonic and depends on the complex interplay between optimizer dynamics, circuit depth, and the specific regions of parameter space explored. Due to the stochastic nature of Differential Evolution and the non-convex QAOA landscape, no single discretization consistently outperforms others across all circuit depths. For $p=2$, medium discretization achieves the highest hit rate ($27.6\%$), while for $p=3$, coarse discretization performs best ($13.5\%$), and for $p=4$, fine discretization unexpectedly surpasses medium ($1.3\%$ vs $0.5\%$). These variations highlight that cache effectiveness is driven by the natural behavior of the optimization process rather than by a simple function of discretization granularity. Nevertheless, coarse discretization generally maximizes reuse by collapsing larger regions of parameter space into fewer equivalence classes, at the cost of reduced search resolution.

\begin{table}[t]
\small
\centering
\caption{Final cache statistics after the full DE run (population 500, 50 iterations).}
\label{tab:qaoa_stats}
\begin{tabular}{lccccc}
\toprule
$p$ & Discretization & Calls & Hits & Hit rate & Cache entries \\
\midrule
2 & Coarse & 25,500 & 2,562 & 10.05\% & 22,530 \\
2 & Medium & 25,500 & 7,044 & 27.62\% & 18,155 \\
2 & Fine & 25,500 & 2,228 & 8.74\% & 13,189 \\
3 & Coarse & 25,500 & 3,446 & 13.51\% & 22,892 \\
3 & Medium & 25,500 & 1,573 & 6.17\% & 23,204 \\
3 & Fine & 25,500 & 744 & 2.92\% & 16,173 \\
4 & Coarse & 25,500 & 918 & 3.60\% & 24,063 \\
4 & Medium & 25,500 & 126 & 0.49\% & 19,692 \\
4 & Fine & 25,500 & 322 & 1.26\% & 19,813 \\
\bottomrule
\end{tabular}
\end{table}

\begin{figure}[htbp]
    \centering
    \includegraphics[width=0.85\linewidth]{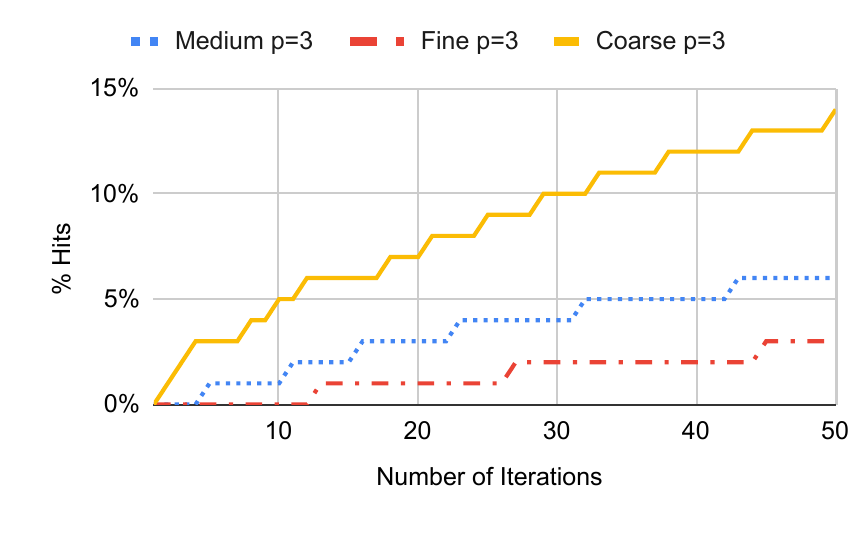}
    \caption{Detailed view for $p=3$ showing cache hit percentage during optimization. Coarser discretizations produce significantly higher reuse throughout the optimization trajectory.}
    \label{fig:qaoa_tradeoff}
\end{figure}

Figure~\ref{fig:qaoa_tradeoff} illustrates the trade-off between discretization resolution and cache efficiency. Rather than representing a limitation, this trade-off exposes a co-design opportunity between algorithm configuration and systems efficiency: discretization simultaneously controls optimization granularity and computational reuse.

\begin{figure}
    \centering
    \includegraphics[width=\linewidth]{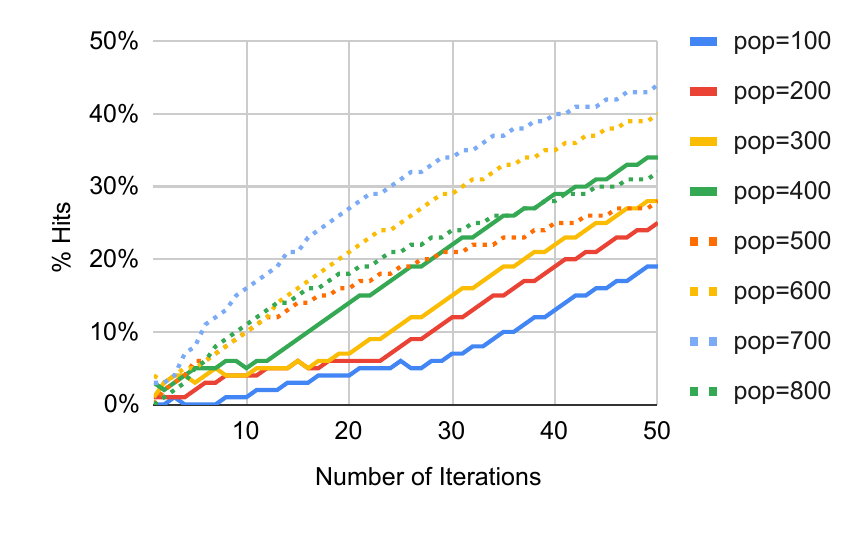}
    \caption{Avoided circuit simulations as a function of DE population size. Larger populations increase concurrency and amplify the effectiveness of semantic-aware circuit caching.}
    \label{fig:qaoa_scalability}
\end{figure}

Finally, Figure~\ref{fig:qaoa_scalability} evaluates scalability with respect to population size by measuring the number of avoided circuit simulations as DE population increases. Larger populations generate higher concurrent request rates to the cache, amplifying reuse opportunities and increasing the absolute number of avoided simulations. This result highlights that circuit caching becomes increasingly beneficial as hybrid quantum workflows scale toward HPC-sized parallelism.

Overall, this evaluation demonstrates that a substantial fraction of QAOA optimization effort is redundant once circuit equivalence is taken into account. By identifying and caching ZX-equivalent circuits, distributed variational optimization eliminates repeated execution of semantically identical quantum computations without modifying the optimizer or execution backend. These results position semantic circuit caching as a versatile, backend-independent systems optimization applicable across a broad class of hybrid quantum-classical workloads.

\begin{table*}[htbp]
\small
\centering
\caption{Storage growth for LMDB and Redis backends. Full results store complete statevectors (wire cutting), while compact storage retains only expectation values (QAOA).}
\label{tab:cache_growth}
\begin{tabular}{lcccc}
\toprule
Num Circuits & LMDB Full & LMDB Compact & Redis Full & Redis Compact \\
\midrule
100     & 4.81 MB   & 0.08 MB   & 24.44 MB  & 15.36 MB \\
500     & 23.58 MB  & 0.10 MB   & 122.22 MB & 76.80 MB \\
1,000   & 47.04 MB  & 0.12 MB   & 244.44 MB & 153.60 MB \\
5,000   & 234.74 MB & 0.32 MB   & 1.22 GB   & 0.77 GB \\
10,000  & 469.37 MB & 0.57 MB   & 2.44 GB   & 1.54 GB \\
100,000 & 4.69 GB   & 5.70 MB   & 24.44 GB  & 15.36 GB \\
1,000,000 & 46.94 GB & 57 MB & 244.44 GB & 153.60 GB \\
\bottomrule
\end{tabular}
\par\vspace{4pt}
\footnotesize Note: LMDB stores raw binary data with minimal overhead. Redis stores serialized objects plus internal data structure overhead.
\end{table*}

Table~\ref{tab:cache_growth} illustrates the storage growth for both LMDB and Redis backends. LMDB stores raw binary data directly in a memory-mapped file with minimal per-entry overhead (16-32 bytes), resulting in approximately 480 bytes per full statevector and 24 bytes per compact expectation value. Redis, in contrast, stores serialized objects plus internal data structure overhead (dictionary entries, pointers, reference counting, and memory fragmentation). Redis requires 0.244 MB (244 KB) per full statevector and 0.154 MB (154 KB) per compact value. The difference of 90 KB per circuit corresponds to the statevector data itself, while the compact footprint of 154 KB reflects Redis's per-entry cost, which dominates for small values.

In the wire cutting example, full results (complete statevectors) were stored, representing the largest possible data footprint per circuit. For 10,000 cached circuits, LMDB requires 469 MB, while Redis requires approximately 2.44 GB. In the QAOA example, only compact summaries (expectation values) were stored, reducing the data footprint significantly. However, Redis's per-entry overhead dominates, resulting in 1.54 GB for 10,000 entries, compared to only 0.57 MB for LMDB. Extrapolating to 1,000,000 compact entries, LMDB requires only 57 MB, a remarkably small footprint that makes long-term caching of expectation values highly practical even on resource-constrained systems.

This substantial memory overhead for Redis is a deliberate trade-off: Redis provides superior scalability under high concurrency (as shown in Figures~\ref{fig:hea_runtime} and~\ref{fig:random_runtime}) at the cost of increased memory usage, while LMDB offers memory efficiency and simplicity for lightweight or single-writer deployments. Importantly, Redis distributes its memory footprint across multiple nodes in a cluster, so the per-node memory burden remains modest even for large caches. In contrast, LMDB stores the entire cache in a single file, which can become a bottleneck for distributed access. Additionally, LMDB is inherently more portable, as the entire cache resides in a single file that can be easily copied, shared, or archived. To leverage the strengths of both backends, our cache architecture supports cross-backend persistence: a Redis cluster can be exported to an LMDB database for long-term storage or portability, and an LMDB cache can be imported into Redis when high-concurrency access is required. Although not explicitly evaluated in this paper, this conversion capability provides flexibility for hybrid workflows that transition between deployment environments.

%% file: Sections/6_Discussion.tex
\section{Discussion}
\label{sec:discussion}

While our evaluation focused on wire cutting and QAOA optimization, semantic circuit caching generalizes naturally to other hybrid quantum-classical workflows. Variational quantum chemistry algorithms, such as UCC and adaptive ans\"atze, generate families of circuits that differ mainly in parameter values or gate orderings, leading to repeated evaluation of structurally equivalent subcircuits that our cache can identify \cite{grimsley2019adaptive}. Similarly, error mitigation techniques like zero-noise extrapolation and probabilistic error cancellation produce multiple circuit variants that implement equivalent logical operations under modified noise conditions; once noise-scaling transformations are abstracted, these circuits share identical computational structure, enabling semantic reuse \cite{temme2017error}.

Our current implementation operates on static circuits, and dynamic circuit execution with classical control and measurement-based branching is not yet supported. However, the semantic representation can naturally handle subcircuits between measurements, and full support for adaptive workflows remains future work \cite{rydberg2023dynamic}. Regarding equivalence detection, ZX-calculus reduction does not guarantee a unique normal form; semantically equivalent circuits may occasionally produce different hashes if reductions converge to different forms, which reduces reuse opportunities but never compromises correctness. Exploring more complete canonicalization or probabilistic equivalence checking is a promising direction. While Weisfeiler--Leman hashing is not guaranteed collision-free, observation across tens of thousands of circuits in our evaluation revealed no false positives or false negatives, and the practical impact of any undetected collision would be limited to an unnecessary cache miss.

While our QPU validation on MareNostrum~Ona (35-qubit superconducting) used limited circuit sizes due to hardware availability, the observed $11.2\times$ speedup confirms the cache's effectiveness on real quantum processors. Despite these limitations, semantic circuit caching provides a practical, scalable systems-level optimization for current and near-term hybrid quantum-classical workloads.

%% file: Sections/7_Conclusions.tex
\section{Conclusion}
\label{sec:conclusion}

In this work, we introduced semantic circuit caching, a systems-level optimization that identifies and reuses previously executed quantum circuits based on semantic equivalence rather than syntactic identity. Our approach combines ZX-calculus circuit simplification with Weisfeiler--Lehman graph hashing to construct reduced circuit representations with hashes that enable efficient detection of equivalent computations across distributed executions. This work contributes to the Qdislib \cite{10.1145/3731599.3767547}, which provides reusable systems-level components for distributed hybrid quantum-classical computing.

We evaluated the proposed approach across multiple hybrid workloads. In distributed wire-cutting experiments, caching eliminated up to 91.98\% of redundant subcircuit simulations, achieving speedups exceeding $7\times$ on a single node and maintaining benefits at scale. In QAOA optimization with Differential Evolution, caching avoided up to 27.6\% of circuit evaluations without altering optimizer behavior or solution quality. Validation on quantum hardware confirmed a $11.2\times$ speedup, demonstrating practical impact beyond simulation.

A key result is that circuit equivalence arises as a structural property of hybrid quantum workflows rather than a workload-specific artifact. Cache effectiveness was largely independent of the storage backend, reinforcing the cache as a reusable systems component. This work suggests a shift toward treating quantum circuits as reusable computational artifacts within distributed infrastructures, like memoization in classical HPC.

Future work includes tighter integration with runtime scheduling, adaptive cache-aware workload placement, heterogeneous execution fully combining CPUs, GPUs, and QPUs, and combining caching with compiler-level optimizations.

Overall, our results demonstrate that exploiting semantic equivalence between quantum circuits substantially reduces redundant computation and improves scalability in hybrid quantum--classical systems. As quantum computing integrates with HPC infrastructures, systems-level techniques such as circuit caching will play a central role in enabling efficient large-scale quantum applications.

%% file: bibliography.bib
@article{backens2014zx,
  title     = {The {ZX}-Calculus is Complete for Stabilizer Quantum Mechanics},
  author    = {Backens, Miriam},
  journal   = {New Journal of Physics},
  volume    = {16},
  number    = {9},
  pages     = {093021},
  year      = {2014}
}

@article{kissinger2020pyzx,
  title     = {{PyZX}: Large Scale Automated Diagrammatic Reasoning},
  author    = {Kissinger, Aleks and van de Wetering, John},
  journal   = {Electronic Proceedings in Theoretical Computer Science},
  volume    = {318},
  pages     = {229--241},
  year      = {2020}
}

@article{duncan2020graphical,
   title={Graph-theoretic Simplification of Quantum Circuits with the ZX-calculus},
   volume={4},
   ISSN={2521-327X},
   url={http://dx.doi.org/10.22331/q-2020-06-04-279},
   DOI={10.22331/q-2020-06-04-279},
   journal={Quantum},
   publisher={Verein zur Forderung des Open Access Publizierens in den Quantenwissenschaften},
   author={Duncan, Ross and Kissinger, Aleks and Perdrix, Simon and van de Wetering, John},
   year={2020},
   month=June, pages={279} }

@misc{li2022quantummultiplevalueddecisiondiagrams,
      title={Quantum Multiple-Valued Decision Diagrams with Linear Transformations}, 
      author={Yonghong Li and Hao Miao},
      year={2022},
      eprint={2207.11395},
      archivePrefix={arXiv},
      primaryClass={quant-ph},
      url={https://arxiv.org/abs/2207.11395}, 
}

@ARTICLE{7163590,
  author={Niemann, Philipp and Wille, Robert and Miller, David Michael and Thornton, Mitchell A. and Drechsler, Rolf},
  journal={IEEE Transactions on Computer-Aided Design of Integrated Circuits and Systems}, 
  title={QMDDs: Efficient Quantum Function Representation and Manipulation}, 
  year={2016},
  volume={35},
  number={1},
  pages={86-99},
  doi={10.1109/TCAD.2015.2459034}}

@article{farhi2014qaoa,
  title={A Quantum Approximate Optimization Algorithm},
  author={Farhi, Edward and Goldstone, Jeffrey and Gutmann, Sam},
  journal={arXiv preprint arXiv:1411.4028},
  year={2014}
}

@article{ostaszewski2019evqe,
  title={Structure optimization for parameterized quantum circuits},
  author={Ostaszewski, Mateusz and Grant, Edward and Benedetti, Marcello},
  journal={Quantum},
  volume={3},
  pages={181},
  year={2019}
}

@article{failde2023de_vqa,
  title   = {Using Differential Evolution to avoid local minima in Variational Quantum Algorithms},
  author  = {Fa{\'i}lde, David and Viqueira, Jos{\'e} D. and Mussa Juane, Manuel and others},
  journal = {Scientific Reports},
  volume  = {13},
  pages   = {16230},
  year    = {2023},
  doi     = {10.1038/s41598-023-43404-3}
}

@article{Peham2022Equivalence,
title={Equivalence Checking of Parameterized Quantum Circuits: Verifying the Compilation of Variational Quantum Algorithms},
author={Tom Peham and Lukas Burgholzer and R. Wille},
journal={2023 28th Asia and South Pacific Design Automation Conference (ASP-DAC)},
year={2022},
pages={702-708},
doi={10.1145/3566097.3567932}
}

@article{Burgholzer2020Advanced,
title={Advanced Equivalence Checking for Quantum Circuits},
author={Lukas Burgholzer and R. Wille},
journal={IEEE Transactions on Computer-Aided Design of Integrated Circuits and Systems},
year={2020},
volume={40},
pages={1810-1824},
doi={10.1109/tcad.2020.3032630}
}

@inproceedings{NetworkX,
  author    = {Aric A. Hagberg and Daniel A. Schult and Pieter J. Swart},
  title     = {Exploring Network Structure, Dynamics, and Function using NetworkX},
  booktitle = {Proceedings of the 7th Python in Science Conference},
  pages     = {11 - 15},
  address   = {Pasadena, CA USA},
  year      = {2008},
  editor    = {Ga\"el Varoquaux and Travis Vaught and Jarrod Millman},
}

@article{LMDB,
  author     = {Henry, Gavin},
  title      = {Howard Chu on Lightning Memory-Mapped Database},
  year       = {2019},
  issue_date = {Nov.-Dec. 2019},
  publisher  = {IEEE Computer Society Press},
  address    = {Washington, DC, USA},
  volume     = {36},
  number     = {6},
  issn       = {0740-7459},
  url        = {https://doi.org/10.1109/MS.2019.2936273},
  doi        = {10.1109/MS.2019.2936273},
  journal    = {IEEE Softw.},
  month      = nov,
  pages      = {83–87},
  numpages   = {5}
}

@misc{Web:redis,
  author       = {{Redis}},
  howpublished = {Web page at \url{https://redis.io/}},
  year         = {(Date of last access: 31th March, 2026)},
}

@misc{Web:redis_cluster,
  author       = {{Redis Cluster Specification}},
  howpublished = {Web page at \url{https://redis.io/docs/latest/operate/oss_and_stack/reference/cluster-spec/}},
  year         = {(Date of last access: 31th March, 2026)},
}

@article{grimsley2019adaptive,
    title = {An adaptive variational algorithm for exact molecular simulations on a quantum computer},
    author = {Grimsley, Harper R and Economou, Sophia E and Barnes, Edwin and Mayhall, Nicholas J},
    journal = {Nature Communications},
    volume = {10},
    number = {1},
    pages = {3007},
    year = {2019},
    publisher = {Nature Publishing Group}
}

@article{temme2017error,
    title = {Error mitigation for short-depth quantum circuits},
    author = {Temme, Kristan and Bravyi, Sergey and Gambetta, Jay M},
    journal = {Physical Review Letters},
    volume = {119},
    number = {18},
    pages = {180509},
    year = {2017},
    publisher = {APS}
}

@inproceedings{rydberg2023dynamic,
author = {Hong, Xin and Feng, Yuan and Li, Sanjiang and Ying, Mingsheng},
title = {Equivalence Checking of Dynamic Quantum Circuits},
year = {2022},
isbn = {9781450392174},
publisher = {Association for Computing Machinery},
address = {New York, NY, USA},
url = {https://doi.org/10.1145/3508352.3549479},
doi = {10.1145/3508352.3549479},
booktitle = {Proceedings of the 41st IEEE/ACM International Conference on Computer-Aided Design},
articleno = {127},
numpages = {8},
location = {San Diego, California},
series = {ICCAD '22}
}

@article{vandeWetering:2020giq,
    author = "van de Wetering, John",
    title = "{ZX-calculus for the working quantum computer scientist}",
    eprint = "2012.13966",
    archivePrefix = "arXiv",
    primaryClass = "quant-ph",
    month = "12",
    year = "2020"
}

@inproceedings{10.1145/3731599.3767547,
author = {Tejedor, Mar and Casas, Berta and Conejero, Javier and Cervera-Lierta, Alba and Badia, Rosa M.},
title = {Orchestrating Quantum-HPC Workflows with Distributed Quantum Circuit Cutting},
year = {2025},
isbn = {9798400718717},
publisher = {Association for Computing Machinery},
address = {New York, NY, USA},
url = {https://doi.org/10.1145/3731599.3767547},
doi = {10.1145/3731599.3767547},
booktitle = {Proceedings of the SC '25 Workshops of the International Conference for High Performance Computing, Networking, Storage and Analysis},
pages = {1898–1906},
numpages = {9},
location = {
},
series = {SC Workshops '25}
}

@article{doi:10.1142/S0219749905001067,
author = {Janzing, Dominik and Wocjan, Pawel and Beth, Thomas},
title = {{"Non-identity-check" is QMA-complete}},
journal = {International Journal of Quantum Information},
volume = {03},
number = {03},
pages = {463-473},
year = {2005},
doi = {10.1142/S0219749905001067},
URL = {https://doi.org/10.1142/S0219749905001067},
eprint = {https://doi.org/10.1142/S0219749905001067}
}

@article{10.1145/3729229,
author = {Kole, Abhoy and Djeridane, Mohammed Elkacem and Weingarten, Lennart and Datta, Kamalika and Drechsler, Rolf},
title = {qSAT: Design of an Efficient Quantum Satisfiability Solver for Hardware Equivalence Checking},
year = {2025},
issue_date = {April 2025},
publisher = {Association for Computing Machinery},
address = {New York, NY, USA},
volume = {21},
number = {2},
issn = {1550-4832},
url = {https://doi.org/10.1145/3729229},
doi = {10.1145/3729229},
journal = {J. Emerg. Technol. Comput. Syst.},
month = jul,
articleno = {6},
numpages = {15}
}

@article{PhysRevLett.125.150504,
  title = {Simulating Large Quantum Circuits on a Small Quantum Computer},
  author = {Peng, Tianyi and Harrow, Aram W. and Ozols, Maris and Wu, Xiaodi},
  journal = {Phys. Rev. Lett.},
  volume = {125},
  issue = {15},
  pages = {150504},
  numpages = {6},
  year = {2020},
  month = {Oct},
  publisher = {American Physical Society},
  doi = {10.1103/PhysRevLett.125.150504},
  url = {https://link.aps.org/doi/10.1103/PhysRevLett.125.150504}
}

@article{WL,
    title = {The reduction of a graph to canonical form and the algebra which appears therein},
    author = {Weisfeiler, Boris and Leman, Andrei},
    journal = {Nauchno-Technicheskaya Informatsiya},
    volume = {2},
    number = {9},
    pages = {12--16},
    year = {1968},
    note = {English translation by Grigory Ryabov available at \url{https://www.iti.zcu.cz/wl2018/pdf/wl_paper_translation.pdf}}
}

@software{qibochem,
  author       = {Mak, Adrian and Le, Tan and Carrazza, Stefano and Candido, Alessandro and Jun, Ye},
  title        = {qiboteam/qibochem: Qibochem 0.0.1 (v0.0.1)},
  year         = {2024},
  publisher    = {Zenodo},
  doi          = {10.5281/zenodo.10473173},
  url          = {https://doi.org/10.5281/zenodo.10473173}
}
